\documentclass[11pt]{article}
\usepackage[textwidth=15.2cm,textheight=22cm]{geometry}
\usepackage{amsmath,amssymb}
\usepackage{latexsym}
\usepackage{multicol}
\usepackage{graphicx}
\usepackage{caption}
\usepackage{cancel}
\usepackage{bm}
\numberwithin{equation}{section}
\pdfoutput=1
\usepackage[T1]{fontenc}
\usepackage[latin9]{inputenc}
\usepackage[active]{srcltx}
\usepackage{esint}
\usepackage{cancel}
\usepackage{color}
\usepackage{ulem}
\allowdisplaybreaks[1]

\newcommand{\del}{\partial}
\newcommand{\be}{\begin{equation}}
\newcommand{\ee}{\end{equation}}
\newcommand{\bea}{\begin{eqnarray}}
\newcommand{\eea}{\end{eqnarray}}

\newcommand{\nn}{\nonumber}

\newcommand{\ie}{{\it i.e.\ }}

\newcommand{\ndt}{\noindent}

\newcommand{\parm}{\partial_-}

\newcommand\fr[1]{\frac{1}{#1}}

\definecolor{ggreen}{rgb}{0.0, 0.5, 0.20}

\begin{document}

	\begin{titlepage}
		\begin{flushright}    
			{\small $\,$}
		\end{flushright}
		\vskip 1cm
		\centerline{\Large{\bf{Bondi-Metzner-Sachs algebra as an extension of the }}}
		\vskip .5cm
		\centerline{\Large{\bf{Poincar\'e symmetry in light-cone gravity}}}
		\vskip 1.5cm
		\centerline{Sudarshan Ananth$^\dagger$, Lars Brink$^*$ and Sucheta Majumdar$^\ddagger$}
		\vskip .7cm
		\centerline{$\dagger$\,\it {Indian Institute of Science Education and Research}}
		\centerline{\it {Pune 411008, India}}
		\vskip 0.7cm
		\centerline{$^*\,$\it {Department of  Physics, Chalmers University of Technology}}
		\centerline{\it {S-41296 G\"oteborg, Sweden}}
		\vskip 0.7cm           
		\centerline{$\ddagger$\,\it{Universit\'e Libre de Bruxelles and International Solvay Institutes,}}
		\centerline{\it{Campus Plaine CP231, B-1050 Brussels, Belgium}}
		\vskip 1.5cm
		\centerline{\bf {Abstract}}
		\vskip .5cm
		
		\ndt We analyze possible local extensions of the Poincar\'e symmetry in light-cone gravity in four dimensions. We use a formalism where we represent the algebra on the two physical degrees of freedom, one with   {helicity}  $2$ and the other with helicity $-2$. The representation is non-linearly realized and one of the light-cone momenta is the Hamiltonian, which is hence a non-linear generator of the algebra. We find that this can be locally realized and the Poincar\'e algebra extended to the BMS symmetry without any reference to asymptotic limits.

		\vfill
	\end{titlepage}

	\section{Introduction}
	
	\ndt		
	In a series of papers, over the years, we have studied gravity and Yang-Mills theories in a formalism we have named the $lc_2$ formalism. In this framework, the (super)Poincar\'e algebra is spanned on just the physical degrees of freedom and one of the light-cone directions is identified as the evolution (``time'') variable. By doing this, the conjugate light-cone momentum is the Hamiltonian and can be read off by constructing the generators of the algebra order by order in the coupling constant~\cite{Bengtsson:1983pd}. The formalism for gravity is set up on a Minkowski background and the price to pay is that the Hamiltonian is an infinite series of terms which after the four-point coupling becomes unmanageable. This means that we never fully prove the symmetries we realize but in our experience, checking symmetries up to the three-point coupling is sufficient to be convinced that the theory in question satisfies that particular symmetry. (We do know them all to the four-point coupling order.) Since this is a perturbative expansion in the gravity fields we can only trust it for weak enough fields.
	\vskip 0.2cm
	\ndt
	In a fairly recent paper we showed that (within the limits alluded to above) the famous Ehlers symmetry~\cite{Ananth:2018hif,Majumdar:2019vic} can be  shown to be an additional symmetry, even in the four-dimensional theory. In this letter, we ask a different question. Can any of the Poincar\'e generators be lifted to be local? We will find that the Hamiltonian can indeed be made local and we can identify it with the Bondi-van der Burg-Metzner-Sachs (BMS) group  {which} is an infinite-dimensional enhancement of the Poincar\'e group.  In the usual formulation, this arises as the asymptotic symmetry group at null infinity for asymptotically flat spacetimes~\cite{Bondi:1962px,Sachs,Sachs:1962zza}. There has been a renewed interest in the study of asymptotic symmetries following a recent body of work~\cite{Strominger:2017zoo}, wherein these symmetries have been connected to soft theorems for gauge theories. Here we will show the symmetry as a symmetry in the bulk avoiding the usual sensitivity to the boundary conditions and gauge choices imposed on the fields. At spatial infinity, for instance, under the standard boundary conditions in the Hamiltonian formulation of gravity, only the Poincar\'e algebra could be canonically realized, not the BMS algebra~\cite{Regge:1974zd}. In a recent work, however, the BMS group was recovered at spatial infinity by suitably relaxing the  boundary conditions, thereby, resolving a longstanding disparity between the asymptotic structure at spatial and null infinity~\cite{Henneaux:2018cst,Henneaux:2019yax}.
	\vskip 0.2cm
	\ndt	
	In the next section we will review shortly the formalism and in the following ones we will show the invariance of the Hamiltonian under the extended symmetry and connect the symmetry to the BMS one. In a subsequent paper we will discuss the problem in more generality relating the symmetry to residual reparameterization invariance and investigate if the symmetry can be further extended.

	 	\section{Light-cone gravity and Poincar\'e symmetry}
	
	\ndt With the metric $(-,+,+,+)$, the light-cone coordinates are defined as
	\bea \label{coord}
	x^\pm=\fr{\sqrt 2}(x^0\pm x^3)\ ,&& x=\frac{1}{\sqrt 2}\,(\,{x_1}\,+\,i\,{x_2}\,)\ , \qquad \bar x= x^*\,,
	\eea
	with $\partial_\pm, \bar\partial, \partial$ being the corresponding derivatives. 
	
	\vskip 0.2cm
	\ndt The Poincar\'e algebra is spanned on the two physical degrees of freedom which we call $h(x)$ with helicity $2$ and $\bar h(x)$ with helicity $-2$.   {We have chosen $x^+$ as the evolution parameter, ``time'' according to Dirac's prescription~\cite{Dirac}. One of the Poincar\'e generators, the conjugate momentum $P_+$, is the light-cone Hamiltonian. The light-cone Poincar\'e generators split into two kinds: the kinematical ones ($\mathbb K$) which do not involve the time derivative $\del_{+}$ and the dynamical ones that involve time derivatives and hence get non-linear contributions in the interacting theory~\cite{Bengtsson:1983pd}. Accordingly, we have the generators (the missing indices refer to the transverse direction as in (\ref{coord}))
	\bea \label{K-D split}
	&&\mathbb K:\quad \{P, \bar P, P_-, J, J^+, \bar J^+, J^{+-}\}\,,  \nn \\
	&& \mathbb D:\quad \{P_+\equiv H, J^-, \bar J^-\}\,.
	\eea
	\ndt 
	and the corresponding algebraic structure	
	\bea \label{KDcomm}
	[\, \mathbb K, \mathbb K\,]~=~ \mathbb K\, ,& [\,\mathbb K, \mathbb D\,]~=~ \mathbb D\,, & [\mathbb D , \mathbb D\,]~=~ 0\, .
	\eea
For completeness, we present the light-cone Poincar\'e algebra in the appendix. The canonical generators of the light-cone Poincar\'e algebra for gravity in terms of the physical fields $h,\ \bar h$ can be found in~\cite{Bengtsson:1983pd}.
\subsection*{Gravity in light-cone gauge}
\vskip 0.2cm
	\ndt The Hamiltonian for gravity reads~\cite{Bengtsson:1983pd,Ananth:2006fh}
	\bea \label{LCH}
	H &=& \int d^3 x\, \left\{\del \bar h\, \bar \del h\ +\ 2\, \kappa\, \del_-^2 \bar h\, \left(h\,\frac{{\bar \del}^2}{\del_-^2}h\ -\ \frac{\bar \del}{\del_-}h\, \frac{\bar \del}{\del_-}h\right) + c.c.\ +\ \mathcal O(\kappa^2)\right\}\,.
	\eea
	The inverse derivative $\frac{1}{\parm}$, defined by $\frac{1}{\parm} (\parm h) = h$, introduces a mild non-locality along the light-cone which is easy to handle~\cite{Mandelstam:1982cb}. An alternative is to Fourier transform to momentum space leading to a new kind of pole.
	\vskip 0.2cm
	\ndt Given the following relation
\bea
\delta_H h\,\equiv\,\{h\,,\,H\}\ ,
\eea
where the r.h.s. is the Poisson bracket of the field with the Hamiltonian\footnote{The fields $h$ and $\bar{h}$ satisfy the relations \[\{h(x), \bar{h}(y)\}\, =\, \frac{1}{\del_-} \delta^{(3)} (x-y)\, ,\quad \{\, h(x), h(y)\,\}\, =\, \{\,\bar h(x), \bar{h}(y)\,\}\, =\, 0\, .\]}, we obtain 
	\bea \label{delta}
	\delta_H h&=& \frac{\del \bar \del}{\del_-}h\, +\, 2\,\kappa\,\del_- \left(h\, \frac{{\bar \del}^2}{\del_-^2} h\, -\, \frac{\bar \del}{\del_-}h\, \frac{\bar \del}{\del_-} h\right)\, \nn \\
	&&+\,2\,\kappa\, \frac{1}{\del_-^3} 
	\left(\frac{\del^2}{\del_-^2}\bar{h}\,\del_-^2 h\, -\, 2\, \frac{\del}{\del_-}\bar{h}\, \del_- \del h\, +\, \bar{h}\,\del_-^2\del^2h  \right),
	\eea
	with $\delta_H \bar h$ being the complex conjugate of the above expression.

	\vskip 0.2cm
	\ndt
Another interpretation of (\ref {delta}) is as a consequence of a shift in the time coordinate
	\bea
	x^+ & \rightarrow& x^+ \, + a\ ,
	\eea	
	\ndt
	with the infinitesimal constant, $a$ not explicitly shown.

\subsection*{Extension of Poincare symmetry - step 1}

\ndt It is natural to ask what happens if this constant is replaced by a function of spacetime, \ie
	\bea
\label{sm}
	x^+ & \rightarrow& x^+ \, + f(x, \bar x, x^+, x^-)\ .
	\eea
	\ndt
	Under this, the fields transform non-linearly as 
	\bea \label{flocal}
	\delta_{H_f} h&=& f(x, \bar x, x^+, x^-)\ \Bigg\{\,\frac{\del \bar \del}{\del_-}h\, +\, 2\,\kappa\,\del_- \left(h\, \frac{{\bar \del}^2}{\del_-^2} h\, -\, \frac{\bar \del}{\del_-}h\, \frac{\bar \del}{\del_-} h\right)\, \nn \\
	&&+\,2\,\kappa\, \frac{1}{\del_-^3} 
	\left(\frac{\del^2}{\del_-^2}\bar{h}\,\del_-^2 h\, -\, 2\, \frac{\del}{\del_-}\bar{h}\, \del_- \del h\, +\, \bar{h}\,\del_-^2\del^2h\right)\,+\, \mathcal O(\kappa^2) \Bigg\}\,,
	\eea
	with $\delta_{H _f}\bar h$  given by the complex conjugate of the above expression. 
\vskip 0.2cm
	\ndt
	It is now interesting to see under what circumstances, the transformation in (\ref {sm}) represents a symmetry of the Hamiltonian. To this end, we compute
	\bea
	\delta_{H_f} H &=& \delta_{H_f}^{(0)}\, H^{(0)} \,+\, \delta_{H_f}^{(\kappa)}\, H^{(0)}\,+\, \delta_{H_f}^{(0)} \,H^{(\kappa)} \,+\, \mathcal O(\kappa^2)\,.
	\eea
	\vskip 0.2cm
	\ndt
	At order $\kappa^0$, we have
	\bea
	\delta_{H_f}^{(0)}\, H^{(0)}&=& -\,(\del_-f)\  \frac{\del \bar \del}{\del_-}\,\bar h\ \frac{\del \bar \del}{\del_-}\, h\,,
	\eea
	which vanishes if $\del_- f=0$. Thus, the invariance of the free Hamiltonian constrains the parameter $f$ to be $f(x, \bar x, x^+)$.
	\vskip 0.2cm
	\ndt At the cubic order, we find 
	\bea \label{cubic}
	\delta_{H_f}^{(0)} \,H^{(\kappa)} &=&  2\, \kappa\, \frac{1}{\del_-^2} (h\,\del_-^2 \bar h)\, \left( \bar{\del}^2 f\, \frac{\del \bar \del}{\del_-}h\,+\, 2\, \bar{\del} f\, \frac{\del \bar{\del}^2}{\del_-}h \right)\,  -\,4\,\kappa\, \del_-^2 \bar h\, \bar{\del} f\, \frac{\del \bar{\del}}{\del_-^2}h\, \frac{\bar{\del}}{\del_-}h\, \nn \\ [0.2cm]
	&&+\,2\, \kappa\, f\, \del_- \bar \del \del \bar h\, \left(h\, \frac{{\bar \del}^2}{\del_-^2} h \, -\, \frac{\bar \del}{\del_-}h\, \frac{\bar \del}{\del_-} h\right)\, + 2\,\kappa\, f\, \del_-^2 \bar h \left( \frac{\del \bar \del}{\del_-}h\right)\, \frac{{\bar \del}^2}{\del_-^2} h\nn \\ [0.2cm]
	&& +\, 2\, \kappa\, f\,\frac{1}{\del_-^2} (h\,\del_-^2 \bar h)\, \frac{\del \bar{\del}^3}{\del_-}h\,-\, 4\, \kappa\, f \,\del_-^2 \bar h\, \frac{\del \bar{\del}^2}{\del_-^2}h\, \frac{\bar{\del}}{\del_-}h\,. 
	\eea
	
	\ndt
	The second and third line, which involve terms with a free $f$, can be simplified further to cancel exactly against $\delta_{H_f}^{(\kappa)}\, H^{(0)}$. In so doing, we obtain terms involving $\bar{\del}f$ and $\bar{\del}^2f$ which cancel against the first line in (\ref{cubic}). Therefore, the transformation in (\ref{flocal}) represents a symmetry of the light-cone Hamiltonian with the local parameter
	\bea
	f&=& f(x, \bar x, x^+)\, .
	\eea
	It is straightforward to check the commutator of $H_f$ with the other generators of the Poincar\'e algebra. For instance, with the momentum $P$, one finds 
	\bea
	[\,P\,,\, H_f\, ]&=& H_{\hat{f}}\,, \quad \text{with}\ \hat{f}~=~ \del f\, ,
	\eea
 which reduces to the familiar case
 \bea
 [\,P\,,\, H\, ]&=& 0\, ,
 \eea
 when $f$ is a constant.

\subsection*{Extension of Poincare symmetry - step 2}

\ndt Having made an extension to the Poincare symmetry focusing on the time direction, we can now look for similar local extensions in the $x$ and $\bar x$ coordinates
\bea 
x ~ \rightarrow ~ x\ +\ Y(x)\,, && \del_- Y~=~ \del Y~=~ 0\,,\\
\bar x ~\rightarrow ~\bar x\ +\ \overline Y(\bar x)\,, && \del_- \overline{Y}~=~ \bar \del \overline{Y}~=~0\,,
\eea
\ndt
under which the fields transform  as
\bea
\label{rotation}
\delta_{Y, \overline Y}\, h&=& Y(x)\, \bar \del h \,+\, \overline Y(\bar x)\, \del h\, +\, (\del \overline Y - \bar \del Y)\,h \,, \nn \\
\delta_{Y, \overline Y}\, \bar h &=& Y(x)\, \bar \del \bar h\, +\, \overline Y(\bar x)\, \del \bar h\,-\, (\del \overline Y - \bar \del Y)\,\bar h \,.
\eea
\ndt The invariance of the Hamiltonian, however, further restricts the parameters to be
\bea
\label{restrict}
{\bar \del}^2 Y = \del^2 \overline Y= 0.
\eea
reducing them to Lorentz rotations. The above condition constrains $Y$ and $\overline Y$ to be linear in the coordinates, ruling out superrotations as an extension of the Poincar\'e algebra in the bulk (unlike the case at null infinity~\cite{Barnich:2009se}). In terms of the light-cone Poincar\'e algebra, the transformations in (\ref{rotation}) account for the transformations generated by $P, \bar P$ and $J$ in the appendix.

\vskip 0.2cm
\ndt Nevertheless, it is more convenient to use the $Y, \overline Y$ notation for the Poincar\'e transformations in the transverse directions $x$ and $\bar x$ in order to demonstrate how the local transformation in (\ref{flocal}) satisfies non-trivial commutation relations with the ones given in (\ref{rotation}). This allows us to establish connections with the BMS symmetry.
\vskip 0.5cm

	 	\section{Comparison with BMS symmetry}
	\ndt A particularly interesting choice of $f$ involving the rotations $Y$ and $\overline{Y}$, which obey (\ref {restrict}), is 
	\bea \label{fBMS}
	f(x^+,x,\bar x)&=& T (x, \bar x)\ +\ \frac{1}{2}\, x^+\, (\del \overline Y\ +\ \bar \del Y)\, ,
	\eea
	\ndt which coincides with the BMS symmetry~\cite{Barnich:2009se}.
	This choice of $f$ follows from the light-cone gauge conditions imposed on the metric.\footnote{The transformations (\ref{flocal}) and (\ref{rotation}) must respect the light-cone gauge conditions on the metric $g_{\mu\nu}$ in order to ensure that no new degrees of freedom are introduced beyond the $h$ and $\bar h$. In the light-cone gauge, three metric components are set equal to zero~\cite{Scherk:1974zm} \[g_{--}\ =\ g_{-i}\ =\ 0\,,\quad ( i=1,2)\,,\] while the remaining components are parameterized as follows: $\ g_{+-}\ =\ - e^{\phi},\ \ g_{ij}\ =\  e^{\psi} \gamma_{ij} $,
			with $\phi, \psi$ real and $\gamma_{ij}$ real, symmetric and unimodular. The fourth gauge choice is \[\phi = \frac{\psi}{2}\, ,\] fixing the time dependence of $f$ in terms of $Y, \overline Y$.  } Since the light-cone system is a Carroll hypersurface, the above relation can also be obtained from the conformal Carroll transformations~\cite{Duval:2014uva} in the null time direction $x^+$.
	
	\vskip 0.2cm
	\ndt The field $h$ transforms as
	\bea
	\delta_{Y, \overline Y, f}\, h&=& \delta_{Y,\overline{Y}} h\ +\ \delta_f h\, ,
	\eea
	\ndt where the explicit form can be read off from (\ref{flocal}) and (\ref{rotation}). One can choose the initial surface to be $x^+=0$ and evolve the algebra at later times using the dynamical generators, in which case the function $f=T$ in (\ref{fBMS}). Therefore, the BMS transformations in light-cone gravity to order $\kappa$ read
	\bea 
	\delta_{Y, \overline Y, T}\, h&=& Y(x)\, \bar \del h + \overline Y(\bar x)\, \del h\ + (\del \overline Y - \bar \del Y)\,h + T\,\frac{\del \bar \del}{\del_-}\,h  \nn \\
	&&-\, 2\, \kappa\, T\, \del_-\, \left(h\, \frac{{\bar \del}^2}{\del_-^2} h\, -\, \frac{\bar \del}{\del_-}h\, \frac{\bar \del}{\del_-} h\right)  \ -\ 2\, \kappa\, T\, \frac{1}{\del_-}\left(\frac{{\del}^2}{\del_-^2}\bar h\, \del_-^2 h\right)\, \nn \\
	&& -\, 2\, \kappa\, T\, \frac{{\del}^2}{\del_-^3}\, (\bar h\, \del_-^2  h)\, +\ 4\, \kappa\, T\, \frac{\del}{\del_-^2} \left(\frac{\del}{\del_-}\bar h\, \del_-^2 h \right)\, ,
	\eea 
	with the parameters $Y$, $\overline Y$ and $T$ satisfying
	\be \label{cond1}
	\del Y = \del_- Y = 0 \ , \quad \bar  \del \overline Y = \del_- \overline Y = 0\  , \quad \del_- T = 0\ ,
	\ee
	and 
	\be \label{cond2}
	\bar{\del}^2Y~=~\del^2 \overline Y~=~0\, .
	\ee
	The commutator of these transformations is
	\bea \label{alg}
	\left[\, \delta(Y_1, \overline Y_1, T_1)\, ,\ \delta (Y_2, \overline Y_2, T_2) \, \right]\, h &=& \ \delta (Y_{12}, \overline Y_{12}, T_{12})\, h\ ,
	\eea
	\ndt where the new parameters are 
	\bea
	Y_{12}& \equiv& Y_2\, \bar \del\, Y_1\ -\ Y_1\, \bar \del\, Y_2\,, \\
	\overline{Y}_{12} & \equiv & \overline Y_2\, \del\, \overline Y_1\ - \ \overline Y_1\, \del\, \overline Y_2\,, \\
	T_{12} & \equiv & [Y_2\, \bar \del \, T_1\ +\ \overline{Y_2}\, \del\, T_1\ +  \frac{1}{2}\, T_2 (\bar \del Y_1\ +\, \del \overline Y_1) ]\ - \ (1 \leftrightarrow 2)\ .
	\eea
	\ndt Thus, two such transformations close on themselves with the parameters satisfying (\ref{cond1}) and (\ref{cond2}). This is the light-cone realization of the BMS algebra in four dimensions~\cite{Barnich:2009se}.

	\vskip 0.2cm
	\ndt
Following the BMS nomenclature, these ``supertranslations'' labeled by $T(x,\bar x)$ enhance the dynamical part of the Poincar\'e algebra into an infinite-dimensional set, while the kinematical part of the algebra remains unaltered
	\bea \label{k-T split}
	&&\mathbb K~\rightarrow~\mathbb K\,, \nn \\
	&& \mathbb D~\rightarrow~ \mathbb D (T)\,.
	\eea
	\ndt
	The BMS algebra in (\ref{alg}) can then also be written as 
	\bea \label{LC-BMS}
	[\, \mathbb K, \mathbb K\,]~=~ \mathbb K\, ,& [\,\mathbb K, \mathbb D (T)\,]~=~ \mathbb D (T)\,, & [\mathbb D (T) , \mathbb D (T)\,]~=~ 0\,. 	\eea
	\ndt By restricting $T$ to be linear in $x$ or $\bar x$ with the condition $\del^2T= \bar{\del}^2 T=0$, the dynamical part $\mathbb D(T)$ reduces to $\mathbb D$, which corresponds to the Poincar\'e subgroup of the BMS algebra.
	\vskip 0.2cm
	\ndt Unlike in covariant formulations, where the four spacetime translations get enhanced to angle-dependent supertranslations, in the light-cone formalism, \textit{the dynamical part of the Poincar\'e algebra is enhanced to accommodate supertranslations, thereby, yielding the BMS algebra in four dimensions}. In both cases, however, the enhancement of the Poincar\'e algebra to the infinite-dimensional BMS algebra involves only a single local parameter as a consequence of Lorentz invariance.
	\vskip 0.2cm
	\ndt 
	We must emphasize though that {\it the invariance of the Hamiltonian under the BMS transformations is only strictly proven to order $\kappa$}. However, we do not anticipate formal difficulties in extending these results to higher orders, although the explicit calculations can become extremely cumbersome.
			
\vskip 0.3cm

\begin{center}
* ~ * ~ *
\end{center}

	\ndt Even though this analysis has been performed only up to first order in the coupling constant, one can already draw parallels with existing BMS results for the full Einstein theory at null and spatial infinity. The BMS analysis at spatial infinity involves some subtleties that can be better understood in the linearized theory of gravity, for instance, the canonical realization of boost generators~\cite{Fuentealba:2020ghw}. In the same spirit, our perturbative approach in the light-cone formalism offers an unconventional insight into the structure of the BMS symmetry. Another similarity with the spatial infinity analysis is that the invariance of the Hamiltonian reduces the parameters, $Y$ and $\overline Y$, to Lorentz rotations, thereby eliminating superrotations~\cite{Barnich:2009se} from the theory. There are several ways to interpret the BMS symmetry in gravity apart from the original work of~\cite{Bondi:1962px,Sachs,Sachs:1962zza} as the enhancement of the Poincar\'e algebra with angle-dependent supertranslations. In~\cite{Duval:2014uva}, for instance, the BMS algebra was argued to be the conformal extension of the Carroll algebra on null hypersurfaces. In light-cone gravity, we present a new interpretation of the BMS symmetry as the local extension of the dynamical part of the Poincar\'e algebra in Dirac's front form analysis~\cite{Dirac}. Light-cone gravity offers a straightforward approach to study the BMS symmetry in a physical gauge, which might help us better understand its connection with on-shell amplitudes and soft theorems~\cite{Strominger:2017zoo, Donnay:2018neh}. 
	
	\vskip 0.2cm
	\ndt
	A unique result of our light-cone analysis is that the BMS algebra can be represented on the two physical fields of graviton in the bulk, without any need for asymptotic limits, suggesting that the BMS symmetry could be more than {\it {just}} an asymptotic symmetry.
	Our scope is different from the BMS analyses relevant for black hole physics as we are mainly interested in the implications the extended symmetry might have for the quantum theory in a pertubative regime since the formalism is not suitable for studies with strong gravity fields. We wish, for example, to utilize the knowledge of this new symmetry to better understand the ultraviolet behaviour of the theory, classify counterterms, etc. in the regime of linearized gravity. Light-cone gravity (formulated in the helicity basis) is closely related to on-shell physics - MHV Lagrangians, double-copy relations, etc.~\cite{MHV Yang-Mills, Ananth:2007zy}. Accordingly, the light-cone formalism is well suited for examining the connections between these infinite-dimensional symmetries and the perturbative S-matrix and scattering amplitudes. It will also be interesting to look at the corresponding symmetry issues in maximal supergravity.

	\section*{Acknowledgments}
	We thank Glenn Barnich and Marc Henneaux for many helpful discussions. The work of SM is partially supported by the ERC Advanced Grant ``High-Spin-Grav'',  by FNRS-Belgium (conventions FRFC PDRT.1025.14 and  IISN 4.4503.15), as well as by funds from the Solvay Family.

\appendix
\section{Light-cone Poincar\'e algebra in four dimensions}	  

We define 
\be
J^+\ =\ \frac{J^{+1}+iJ^{+2}}{\sqrt{2}}\ ,\quad \bar J^{+}\ =\ \frac{J^{+1}-iJ^{+2}}{\sqrt 2}\ ,\quad J\ =\ J^{12}\ ,\quad H\ =\ P^-\ .
\ee
\vskip 0.2cm
\ndt
The {\it {non-vanishing}} commutators of the Poincar\'e algebra are listed below.
\bea
&[H, J^{+-}]\ =\ -i H \ , \quad  &[H, J^+] = -i P\ , \qquad \qquad  [H, \bar J^+]\ =\ -i \bar P \nn \\[0.2cm]
&[P^+, J^{+-}]\ = \ iP^+\ ,\quad  &[P^+, J^-]\ =\ -iP\ , \qquad \quad [P^+, \bar J^-]\ =\ -i \bar P \nn \\[0.2cm]
&[P, \bar J^-]\ =\ -i H\ ,  \quad &[P, \bar J^+] \ =\ -iP^+\ , \quad \qquad [P, J]\ =\ P \nn \\ [0.2cm]
&[\bar P, J^-]\ = \ -i H\ , \quad &[\bar P, J^+]\ =\ -iP^+\ , \qquad \quad [\bar P, J]\ =\ -\bar P \nn \\ [0.2cm]
&[J^-, J^{+-}]\ =\ -i J^- \ , \quad &[J^-, \bar J^+]\ =\ iJ^{+-} +  J \ , \quad [J^-, J]\ =\ J^- \nn \\ [0.2cm]
&[\bar J^-, J^{+-}]\ = \ -i \bar J^- \ , \quad &[\bar J^-, J^+]\ =\ iJ^{+-} - J \ , \quad [\bar J^-, J]\ =\ -\bar J^- \nn \\ [0.2cm]
&[J^{+-}, J^+]\ =\ -i J^+ \ ,  \quad &[J^{+-}, \bar J^+]\ = \ -i \bar J^+ \ , \nn \\ [0.2cm]
&[J^+, J]\ =\ J^+\ , \quad &[\bar J^+, J]\ =\ -\bar J^+\ .
\eea

\end{document}